# Vers une Substitution des Services Web sans Inconsistance Sémantique

**Boudjemaa Boudaa**

*Département Informatique*
*Université Ibn Khaldoun*
*Tiaret, Algérie*
*boudjemaa.boudaa@univ-tiaret.dz*

ABSTRACT. *In order to ensure high availability of Web services, recently, a new approach was proposed based on the use of communities. In composition, this approach consists in replacing the failed Web service by another web service joining a community offering the same functionality of the service failed. However, this substitution may cause inconsistency in the semantic composition and alter its mediation initially taken to resolve the semantic heterogeneities between Web services. This paper presents a context oriented solution to this problem by forcing the community to adopt the semantic of the failed web service before the substitution in which all inputs and outputs to/from the latter must be converted according to this adopted semantic, avoiding any alteration of a semantic mediation in web service composition.*

KEYWORDS: *Failed Web Service, Substitution, Community, High-Availability, Context, Semantic Mediation.*

RÉSUMÉ. *Dans le but d'assurer la haute disponibilité des services Web, récemment, une nouvelle approche a été proposée basée sur l'utilisation des communautés. Dans une composition, cette approche consiste à substituer le service Web échoué par un autre service Web adhérant à une communauté offrant la même fonctionnalité du service échoué. En revanche, cette substitution peut provoquer une inconsistance sémantique dans la composition et altérer sa médiation initialement prise pour résoudre les hétérogénéités sémantiques entre ses services Web. Ce papier présente une solution orientée contexte à ce problème et qui consiste à forcer la communauté d'adopter la sémantique du service Web échoué avant la substitution où toutes les entrées et sorties à/de cette dernière doivent être converties suivant cette sémantique adoptée, évitant par cela l'altération de toute médiation sémantique prise dans une composition des services Web.*

MOTS-CLÉS : *Service Web Echoué, Substitution, Communauté, Haute-Disponibilité, Contexte, Médiation Sémantique.*

B.Boudaa

**1. Introduction**

Les services Web émergent comme une nouvelle génération des technologies du Web pour faciliter l'interopérabilité des systèmes d'informations en déployant des interactions automatisées entre les applications d'entreprises réparties voire hétérogènes en se fondant sur les standards Internet comme XML et HTTP. Par contre, les challenges scientifiques tels que la disponibilité, la consistance sémantique, la sécurité, l'intégrité et la maintenance restent toujours à relever. Dans ce papier, nous nous intéressons à la consistance sémantique entre les services Web et leur disponibilité qui est un attribut de fiabilité qualifiant la promptitude d'une application (Avizienis, 2004). L'approche proposée par Maamar et al. (Maamar, 2009), basée sur le fonctionnement des communautés des services Web pour soutenir la haute disponibilité de ces derniers a démontré des solutions aux inconvénients posés par les solutions traditionnelles basées sur la réplication. Dans une composition des services Web, cette approche consiste à la substitution du service échoué par un autre de même fonctionnalité appartenant à une communauté des services Web. Cette substitution va garantir la continuité du service offert, mais d'un autre côté et suite à la sémantique particulière de ce service Web substitué, provoquera de nouvelles hétérogénéités sémantiques qui peuvent surgir au niveau de la composition. La résolution de ces hétérogénéités sémantiques est appropriée pour éviter toute inconsistance sémantique produite par la substitution du service Web échoué.

Dans cet article, nous proposons une solution à ces nouvelles hétérogénéités sémantiques issues de la substitution des services Web qui assurent toujours la médiation sémantique dans une composition, et qui s'appuie sur l'adoption de la sémantique du service Web échoué par la communauté des services Web qui le soutient.

Le reste de ce document est organisé comme suit. La section 2 suggère quelques généralités sur les services Web. La haute disponibilité des services Web est discutée dans 3. La section 4 expose, par un exemple, les travaux de médiation sémantique basée sur la notion de contexte. Notre contribution, qui se résume dans la proposition d'un processus de médiation sémantique en cas de substitution d'un service défaillant, est détaillée en 5. A la fin et dans 6, une conclusion pour terminer.

**2. Services Web et Composition**

Un service Web est une application logicielle accessible qui est capable d'être trouvée et invoquée par d'autres applications pour accomplir leurs divers besoins. Le fonctionnement des services Web est décrit suivant l'architecture SOA (Services Oriented Architecture) en se basant sur plusieurs standards tels que SOAP, WSDL et UDDI.

L'apport principal des services Web est leur participation efficace dans des compositions. Une composition consiste à combiner les fonctionnalités de plusieurs services au sein d'un même processus métier dans le but de répondre à des demandes complexes qu'un seul service ne pourrait pas satisfaire. Le langage d'orchestration WS-BPEL permettra de décrire le processus métier d'une composition des services Web.

## 3. Haute-Disponibilité des Services Web

Il est incontournable que les applications basées sur les services Web soient toujours disponibles et leurs données doivent rester accessibles dans un environnement dynamique et ouvert comme celui d'Internet (par exemple, les services qui permettent de contrôler un avion en vol). Divers événements, comme la coupure de courant, aboutissent à l'arrêt des applications. Ce qui exige de prendre des actions correctives afin d'assurer la disponibilité des services.

De ce fait, la disponibilité est la capacité de fournir un certain niveau de service pendant une situation où un ou plusieurs composants d'un système ont échoué. L'échec peut être programmé (maintenance planifiée) ou non-programmé (panne). La réalisation de la haute disponibilité est d'éliminer les seuls points de l'échec se résumant dans l'existence d'une seule instance d'une ressource. L'élimination de l'échec exige inévitablement l'approvisionnement par des ressources additionnelles. Le but est que, quand un échec se produit, les utilisateurs peuvent encore accéder au service. Le portail du commerce électronique « eBay » a perdu 5 millions de dollars en avril 2002 dus à une panne de serveur qui dura 22 heures; aucun plan approprié de rétablissement d'échec n'était prévu (Maamar, 2009).

La disponibilité est mesurée en utilisant un pourcentage composé de '9' (Nines Notation): 99% désigne le fait que le service est **indisponible** moins de 3,65 jours par an.

### 3.1. *Soutien de la haute-disponibilité*

Les solutions traditionnelles réalisant la disponibilité des services Web sont basées sur la réplication (Salas, 2006) qui est la distribution des copies d'une application logicielle (dans notre cas, service Web) dans un réseau selon l'une des stratégies : active, passive ou hybride (Wiesmann, 2000), et si l'une des copies défaille, les autres peuvent continuer à fournir le service. Cette approche connait plusieurs inconvénients à savoir:
− La synchronisation récurrente des états des copies pour assurer la cohérence.
− La mise à jour du code de ces copies quand un changement s'effectue.

Pour apporter une solution à ces inconvénients, Maamar et al. proposent une nouvelle approche pour soutenir la haute disponibilité (Maamar, 2009) basée sur l'architecture des communautés qui collectent les services Web suivant leurs fonctionnalités (Bentahar, 2007; Maamar, 2007). Cette approche exige que le service Web Maître de la communauté lance un appel d'offres à tous ses services Web Esclaves pour remplacer le service défaillant dans une composition. Parmi tous les services Web Esclaves répondant positivement, le plus approprié selon la qualité de service (disponibilité, fiabilité, temps de réponse,…) sera sélectionné pour fournir la même fonctionnalité que celle du service Web échoué (Maamar, 2009). Avec cette nouvelle approche on pourra trouver des solutions aux inconvénients de la réplication, par exemple, la mise à jour du code n'est pas nécessaire puisque ces services Web de la communauté ont déjà des codes différents (Maamar, 2007).

## 4. Médiation sémantique

La médiation sémantique est la stratégie qui résout les conflits sémantiques qui peuvent surgir lors de l'échange de données entre les services Web dans une composition.

### 4.1. *Exemple d'illustration*

En vue d'illustrer l'importance de la médiation sémantique dans une composition, nous prendrons l'exemple de planification de voyage en ligne composé de trois services:
- Un service Web de réservation de billets d'avion « FlightBooking », qui est un service européen traitant les prix en « Euro »( désignée par EUR) avec une TVA incluse d'un taux de « 19,6% », un facteur multiplicateur (FM) de « 1 » et dont la date est de format « jj/mm/aaaa ».
- Un service Web de réservation d'une chambre au niveau d'un hôtel « HotelBooking », qui est un service japonais utilisant le «Yen»(JPY), avec une TVA non incluse de «9.3%», un FM de «1000» et dont le format de date est «aaaa/mm/jj».
- Un service Web « EuroBanking » pour payer les prix des services précédents offerts en « Euro » avec une TVA de «19,6%» et un FM égal à « 1 ».

Le processus métier de cette composition est réalisé par le langage WS-BPEL où plusieurs conflits sémantiques apparaissent pendant l'échange des données de:
- dates du service Web «FlightBooking» au service Web «HotelBooking».
- prix du service Web « HotelBooking » au service Web « EuroBanking »

Ces hétérogénéités sémantiques sont dues aux différents contextes des services Web. Ces contextes doivent être pris en considération lors de l'échange des données entre ces derniers.

Le contexte est « tout élément interne ou externe, relatif à la donnée ou même complètement extérieur, qui est nécessaire à l'interprétation correcte de la donnée » (Mrissa, 2008). Alors, le contexte est un ensemble d'éléments appelés propriétés sémantiques et qui décrivent les divers aspects et caractéristiques sémantiques d'un concept. Dans l'exemple, le facteur multiplicateur et la devise associés à la donnée de prix sont des propriétés sémantiques. Mrissa et al. construisent un modèle de description contextuelle (Mrissa, 2006) et la médiation consiste à convertir les contextes des services Web.

**4.2. *Modèle de description orienté contexte***

L'échange correct des données entre les services Web exige en premier lieu de décrire les données de ces services par une présentation sémantiquement compréhensible que le modèle WSDL actuel s'avère incapable de faire. Le modèle de description contextuel (Mrissa, 2006), est proposé pour ce but.

*1) Principaux concepts:* Ce modèle contextuel consiste à représenter une donnée, avec une prise en charge de sa sémantique, par un objet sémantique utilisant des modifieurs statiques et dynamiques (Mrissa, 2006):

*a) Objet sémantique:* Un objet sémantique est un objet de donnée avec une sémantique explicite décrite par son contexte. Un objet sémantique S est un quadruplet, représenté de la manière suivante : $S = (c,v,t,C)$, où « c » est un concept de l'ontologie de domaine (voir ci-après.), « v » est la valeur de l'objet sémantique, « t » est le type dans lequel cette valeur est décrite et « C » est le contexte de l'objet sémantique. Ce contexte est lui-même constitué d'objets sémantiques appelés modifieurs, car ils appartiennent au contexte.

*b) Modifieurs :* Les modifieurs ont la capacité de changer la signification de l'unique objet sémantique auquel ils sont associés et peuvent être de type statique ou dynamique. Un modifieur statique est indépendant des autres modifieurs et a une valeur bien déterminée. Par contre un modifieur dynamique est dependant d'un ou plusieurs autres modifieurs.

*Intégration du contexte dans WSDL :* elle se fait par l'utilisation des ontologies de domaine et des ontologies contextuelles ainsi par l'annotation contextuelle du WSDL (développées ci-dessous) :

*Ontologies de domaine et contextuelles:* La caractéristique majeure du modèle contextuel est l'attachement des ontologies de contexte aux concepts des ontologies de domaine (Mrissa, 2008). Une *ontologie de domaine* contient les concepts les plus généraux d'un domaine de connaissances et leurs relations. Une *ontologie contextuelle* est attachée à chaque concept d'ontologie du domaine pour décrire son contexte.

*Annotation contextuelle de WSDL:* Le métamodèle extensible du WSDL(Mrissa, 2008) permettra l'annotation des parties des messages (paramètres d'entrées/sorties). Cette annotation des services Web a pour but de rendre explicite le contexte des données échangées, de manière à ce que ces dernières puissent être traitées comme des objets sémantiques pour effectuer la médiation sémantique, tout en introduisant les ontologies de domaine et contextuelles.

**4.3** *Médiation sémantique entre les services Web*

Après l'intégration de l'information sémantique dans les descriptions des services Web par ce modèle orienté contexte, nous pouvons construire la médiation entre les services Web pour qu'ils puissent échanger leurs données.  A cet effet, nous distinguons entre l'échange des données dans une communauté (Mrissa, 2008) (entre le client et le service Web Esclave chargé de l'exécution de la requête de ce dernier via le service Web Maître) et celui dans une composition (Mrissa, 2008) (entre les services Web composants) où la médiation est commune entre les deux, en utilisant un même module (Fig. 1) de (Mrissa, 2009).

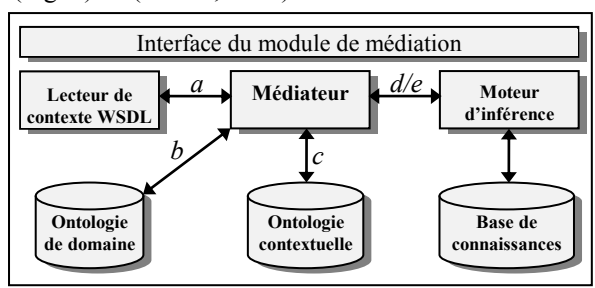

Figure 1.  Module de médiation sémantique

**5. Haute-disponibilité maintenant la médiation**

**5.1.** *Exposition de la problématique*

L'approche de médiation sémantique basée-contexte de Mrissa et al. (Mrissa, 2008), nous permettra de régler les hétérogénéités sémantiques liées aux contextes des services Web de notre exemple. Cette médiation, consiste à ajouter deux services Web médiateurs (Fig. 2) dotés des modules de médiations (Fig.1): un entre « **FlightBooking** » et « **HotelBooking** » pour régler le conflit des dates et l'autre entre « HotelBooking » et « **EuroBanking** » pour régler les prix.

Après un certain temps, la composition qui joue le rôle d'un "Watchdog" (Maamar, 2009) détecte que le service de réservation de billets d'avion « FlightBooking » a cessé de fonctionner et pour assurer la continuation dans son service, l'on a procédé à appliquer l'approche basée-communauté de soutien de la haute disponibilité (voir la section 3). On a fait appel à la communauté nommée « **FlightBooking** » (Fig. 2) qui fournit la même fonctionnalité que celle du service Web échoué. Après le lancement d'un appel d'offre par le Maître « **MasterFlightBooking** », le service Web Esclave « **UKFlightBooking** » a gagné l'offre pour substituer l'échoué. Dans Fig. 2, nous remarquons aussi que le service Web « **EKFlightBooking** » n'a pas bien répondu à l'appel de soutien de son Maître alors que le service Web « **USFlightBooking »** a répondu positivement à l'appel mais n'a pas été choisi par le Maître (à cause de sa qualité de service (QoS) qui est inférieur à celle de « **UKFlightBooking** », par exemple). Or, il peut rester, suivant l'approche, comme support pour le service gagnant (service Primaire) en cas de sa défaillance. « UKFlightBooking » est un service Web anglais qui a ses propres paramètres (la date est de format « mm/jj/aaaa » et le prix est en « livre (Pound) » dont la devise est représentée par GBP avec un facteur multiplicateur de « 1 » et une TVA « non incluse » de « 17.5% »). A ce moment, les résultats retournés (ou requêtes) par ce service substitut doivent être convertis selon le contexte de la communauté. Dans notre cas, le service Web Maître de cette communauté adopte un fichier WSDL annoté avec un contexte Américain différent de celui de « UKFlightBooking » uniquement dans la devise des prix qui est en « **Dollar** » (devise désignée par **USD**). Par la suite, ce service remplaçant rentre en interaction avec le reste des services Web de la composition y compris les services Web médiateurs insérés (avec « HotelBooking » pour le Concept "Date" et avec « EuroBanking » pour le concept "Prix"). Après comparaison des paramètres entrées/sorties de ces services en interaction, l'on a découvert qu'il y a de nouvelles hétérogénéités sémantiques que la médiation actuelle ne peut résoudre avec ses médiateurs déjà intégrés car ils ont des contextes différents de celui du service substitut converti par le contexte de sa communauté (Fig. 2). Il faut donc revoir de nouveau ces actions de médiation initialement prises pour régler ces nouvelles hétérogénéités dues à la substitution. Chose pénible à chaque substitution d'un service Web échoué.

### 5.2. *Substitution et médiation sémantique*

Pour faire face à cette inconsistance sémantique engendrée par la substitution du service Web défaillant, nous présenterons une solution qui consiste, avant le remplacement du service Web défaillant, à faire adopter la sémantique (le contexte) de ce dernier par la communauté (plus précisément par son service Web Maître) soutenant ce service défaillant comme étant sa sémantique. Le service Web Maître va donc procurer le contexte du service défaillant pour annoter son fichier WSDL.

Nous argumentons cette adoption contextuelle par le fait que: (i) le service Web Maître est développé par le concepteur d'application (Mrissa, 2006 ; Maamar, 2007) indépendamment de tout fournisseur de services Web c.-à-d. que le Maître n'a pas de fournisseur qui lui appartient et qui peut réclamer une telle adoption, (ii) Le Maître de la communauté n'implémente pas la fonctionnalité de cette dernière et, par conséquent, ne participe jamais à une composition des services Web. Il est chargé seulement de la gestion de la communauté (Maamar, 2007). Ces caractéristiques permettront au service Web Maître d'adopter un autre contexte autre que le sien sans qu'il y ait une influence sur la médiation sémantique de la composition et (iii) Cette adoption ne change pas les propriétés non fonctionnelles du service Web Maître, elle touche seulement les propriétés fonctionnelles qui sont décrites dans son document WSDL qui fera l'objet de changement. Toute propriété qui caractérise un service et qui n'est pas directement liée à la fonctionnalité délivrée par le service, est considérée comme une propriété non fonctionnelle (Mrissa, 2008).

Avant de faire fonctionner le service remplaçant pour satisfaire la demande de la composition (dans notre cas, c'est **UKFlightBooking** gagnant l'offre), l'on exige au service Web Maître d'adopter la sémantique du service échoué, comme une sémantique de la communauté (Fig. 2). A cet effet, nous avons détaillé un processus de médiation (Boudaa, 2012) composé de deux phases :

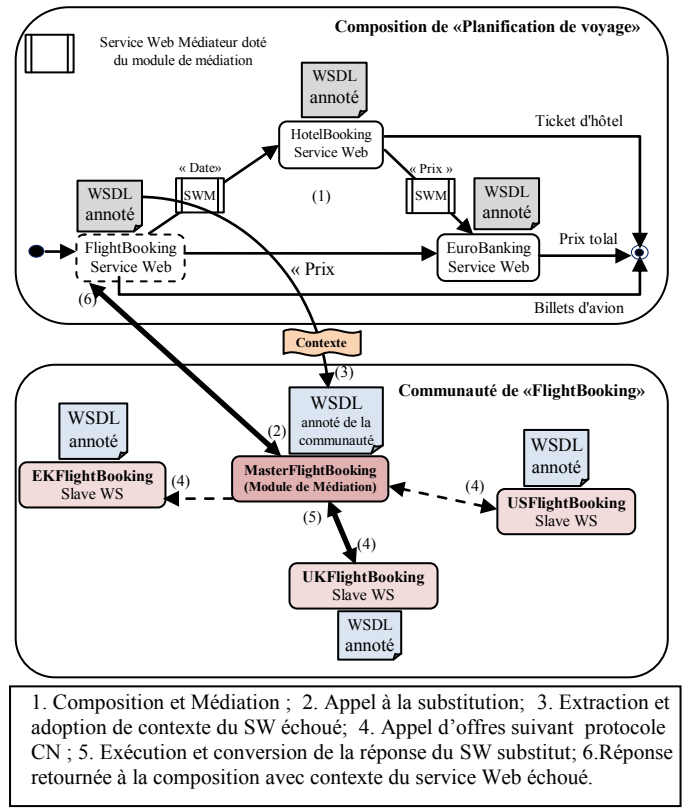

1. Composition et Médiation ; 2. Appel à la substitution; 3. Extraction et adoption de contexte du SW échoué; 4. Appel d'offres suivant protocole CN ; 5. Exécution et conversion de la réponse du SW substitut; 6.Réponse retournée à la composition avec contexte du service Web échoué.

Figure 2.    Vue d'ensemble de la solution proposée

**Extraction et adoption du contexte**. La première phase consiste à extraire la sémantique du service Web défaillant représentée par son contexte et faire l'adopter par le service Web Maître de la communauté appelée pour soutien, et ceci comme suit :

a)   Téléchargement et lecture, par le lecteur de contexte WSDL (Fig. 1) intégré au niveau du service Maître, les fichiers WSDL annotés des services Web Echoué et Maître de la communauté.

b)   Identification du domaine des paramètres d'entrée/sortie des deux services en interrogeant l'ontologie de domaine.

c)   Identification des contextes en communiquant avec l'ontologie contextuelle pour identifier les modifieurs statiques et leurs valeurs contenues dans les annotations. Les modifieurs dynamiques, déduits automatiquement à partir des modifieurs statiques, seront ignorés dans cette première phase chargée, juste, par l'adoption de contexte.

    d) A ce niveau, le module de médiation possède deux objets sémantiques différents dans la mémoire, et il doit extraire, par des méthodes d'extraction, les données contextuelles (nouveau contexte à adopter) du service Web Echoué représentées par les annotations de ses paramètres d'entrées/sorties. Ces données contextuelles seront utilisées dans l'annotation des mêmes paramètres d'entrées/sorties (par correspondance des concepts de l'ontologie de domaine) du service Web Maître tout en écrasant les données contextuelles existantes (ancien contexte à écraser) et cela à l'aide du moteur d'inférence qui s'appuie sur des méthodes d'annotation.

    Après cette étape d'extraction et d'adoption du contexte, la communauté pilotée par son service Web Maître dispose d'un contexte similaire au contexte du service Web échoué.

**Conversion contextuelle**. La deuxième phase, comme citée dans (Mrissa, 2009), est pour faire les conversions sémantiques dans notre médiation et qui est décrite comme suit:

    a) Téléchargement et lecture, par le même lecteur de contexte WSDL, les fichiers WSDL annotés de la communauté et du service Web Esclave choisi.

    b) Identification du domaine des entrées/sorties des deux services en communiquant avec l'ontologie de domaine.

    c) Identification des contextes en utilisant l'ontologie contextuelle pour identifier les modifieurs statiques et leurs valeurs contenues dans les annotations.

    d) Construction des objets sémantiques en déduisant les valeurs des modifieurs dynamiques à partir des valeurs des modifieurs statiques via le moteur d'inférence qui utilise des règles logiques stockées dans la base de connaissance.

    e) A ce stade, le module de médiation possède deux objets sémantiques différents dans la mémoire, et il doit convertir les données du contexte du service Web Esclave au contexte de la communauté en sollicitant encore le moteur d'inférence qui utilise des fonctions de conversions stockées dans la base de connaissance.

    Après cette conversion entre les deux objets sémantiques, la réponse est retournée avec la sémantique adoptée par la communauté, et la composition traite cette réponse comme elle est venue du service Web défaillant. Or, il n'y aura pas de détection de nouveaux conflits sémantiques au niveau de cette composition et par conséquent il n'y aura pas de nécessité de revoir la médiation sémantique et qui reste valable même après la substitution. Dans Fig. 3, nous avons schématisé les étapes de notre processus de médiation par un diagramme de séquence, appliqué à notre scénario, en montrant tous les messages échangés entre les trois acteurs : FlightBooking (service Web échoué), MasterFlightBooking (service Web Maître) et UKFlightBooking (service Web Esclave substitut).

B.Boudaa

La réalisation de notre solution concrétise plusieurs objectifs à la fois, à savoir : (i) éviter d'altérer la médiation sémantique de la composition initialement prise pour régler les hétérogénéités sémantiques entre les services Web participant à la composition, et ce, malgré la substitution du service défaillant par un autre service Web sémantiquement différent, (ii) Permettre le soutien du service Web substitut, en cas de sa défaillance, par les autres services Web Esclaves de support qui ont répondu positivement à l'appel sans aucun changement au niveau de la médiation sémantique de la composition. Même en cas de reprise de son fonctionnement, le service Web échoué peut continuer à fournir sa fonctionnalité dans la composition sans qu'il y ait d'inconvénients et (iii) enfin, notre solution est orientée contexte et ne sort pas du cadre général de l'approche de la médiation sémantique utilisée dans notre composition. Chose qui permettra à notre solution de s'effectuer d'une manière transparente pour le client.

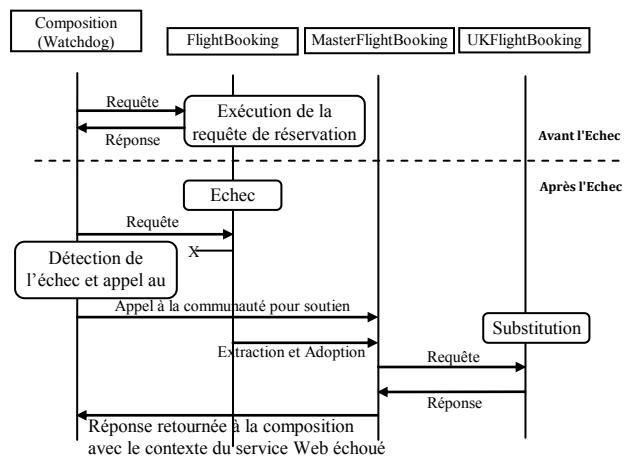

Figure 3. Diagramme de séquence du processus de médiation

### 5.3 *Implémentation*

En vue de prouver la faisabilité de notre solution, l'on a développé une application limitée à notre exemple suivant le diagramme de séquence ci-dessus (Fig. 3).

Dans cette application, deux modules sont majeurs dont le fonctionnement repose sur l'API WSDL4J (http://sourceforge.net/projects/wsdl4j):
1) Le module d'extraction et d'adoption du contexte du service Web échoué.
2) Le module de conversion contextuelle entre les services Web en interaction.

Après la réalisation des étapes pré-requises à savoir ; la construction des ontologies de domaine et contextuelles par l'outil Protégé (http://protege.stanford.edu) et l'annotation des différents services Web par l'éditeur d'annotation de Mrissa et al. (Mrissa, 2008), on effectuera l'extraction et l'adoption du contexte du service échoué par le service Maître par le $1^{er}$ Module et la conversion par le $2^{eme}$ module.

Le listing simplifié suivant illustre cette opération.

*Contexte du service Web échoué «FlightBooking »*
> *<wsdl:part name="Prix_de_ReservationReturn" type="xsd:double" ctxt:context="ctxt1:Price ctxt2:France ctxt2:VATincluded ctxt2:ScaleFactorOne">*

*Contexte du service Web Maître « MasterFlightBooking » (avant l'adoption)*
> *<wsdl:part name="Price_of_ReservationReturn" type="xsd:double" ctxt:context="ctxt1:Price ctxt2:USA ctxt2:VATnotincluded ctxt2:ScaleFactorOne">*

*Contexte du service Web Maître « MasterFlightBooking » (après l'adoption)*
> *{{http://localhost:8080/servlets examples/context.xsd} context= ctxt1:Price ctxt2:France ctxt2:VATincluded ctxt2:ScaleFactorOne}*

La réservation de vol exécutée par le service Web remplaçant doit être retournée par la sémantique du service Web Maître, c.-à-d. par la sémantique adoptée du service Web échoué en évitant tout nouveau conflit peut surgir.

Après déduction des modifieurs dynamiques et leurs valeurs à partir des modifieurs statiques en utilisant les règles logiques, on va construire les objets sémantiques du concept « **Price** » des services Web substitut et Maître pour faire les conversions nécessaires entre leurs contextes à l'aide des fonctions de conversion insérées précédemment dans la base de connaissance du module de médiation. Ces conversions contextuelles sont illustrées dans le listing ci-dessous (en version claire et simplifiée).

> *Prix retourné par le service substitut « UKFlightBooking » :*
> *Price= 1200 (par exemple)*
> *Conversion entre contextes : Anglais - Français*
> *conversion Country : de UK à France #*
> *conversion Currency : de GBP à EUR #*
> *conversion ScaleFactor : de 1 à 1 #*
> *conversion VAT : de False à True #*
> *conversion VATRate : de 17,5 à 19,6 #*
> *Prix retourné la communauté « **FlightBooking** » après les conversions entre les deux contextes: Price= 1575,20.*

Comme travail futur, on prévoit la généralisation de notre application pour qu'elle soit un prototype qui peut agir à une grande échelle sur différentes compositions des services Web.

## 6. Conclusion

Dans ce papier, nous avons présenté une solution orientée contexte pour faire face à l'impact sémantique qui peut être provoqué lors de la substitution d'un service Web échoué pour soutenir sa haute disponibilité dans une composition dotée préalablement par une médiation sémantique. Cet impact surgit à cause de la différence qui peut exister, au niveau sémantique, entre les deux contextes des services défaillant et substitut. La solution détaille un processus de médiation qui évite l'altération de cette médiation sémantique dans la composition. Notre solution permettra à l'approche basée-communauté de soutien de la haute disponibilité des services Web de gagner le challenge de l'influence sémantique surgi par la substitution des services défaillants.

## 7. Bibliographie